# Security Analysis of a Remote User Authentication Scheme with Smart Cards

**Manoj Kumar**

**Department of Mathematics,**

**R.K. (P.G.) College Shamli,**

**Muzaffarnagar, Utter Pradesh- India- 247776.**

E-mail- yamu_balyan@yahoo.co.in

**Abstract**

Yoon et al. proposed a new efficient remote user authentication scheme using smart cards to solve the security problems of *W*. C. Ku and S. M. Chen's scheme. This paper reviews Yoon et al.'s scheme and then proves that the password change phase of Yoon et al's scheme is still insecure. This paper also proves that the Yoon et al. is still vulnerable to parallel session attack.

**Keywords —** Cryptography, Cryptanalysis, Network security, Authentication, Smart cards, Password, Parallel session attack.

## I. INTRODUCTION

To gain the access rights on an *authentication server* (*AS*), a password based remote user authentication schemes is used. The remote user makes a login request with the help of some secret information which are provided by the *AS*. On the other side the *AS* checks the validity of a login request made by a remote user *U***.** In these schemes, the *AS* and the remote user *U* share a secret, which is often called as password. With the knowledge of this password, the remote user *U* uses it to create a valid login request to the *AS***.** *AS* checks the validity of the login request to provide the access rights to the user *U*. Password authentication schemes with smart cards have a long history in the remote user authentication environment. So far different types of password authentication schemes with smarts cards [3,4,5,6,12,13,14,18,20,21,22,23,26,31] have been proposed.

In 1981, Lamport [17] proposed the first well-known remote password authentication scheme using smart cards. In Lamport's scheme, the *AS* stores a password table at the server to check the validity of the login request made by the user. However, high hash overhead and the necessity for password resetting decrease the suitability and practical ability of Lamport's scheme. In addition, the Lamport scheme is vulnerable to a small *n* attack [7]. Since then, many similar schemes [25,28] have been proposed. They all have a common feature: *a verification password table should be securely stored in the AS.* Actually, this property is a disadvantage for the security point of view. Keep in mind all the security requirements for a secure remote user authentication scheme, in 2002, Chien–Jan–Tseng [13] introduced an efficient remote user authentication scheme using smart cards. In 2004, Ku and Chen [33] pointed out some attacks [7,30,32] on Chien – Jan and Tseng's scheme. According to Ku and Chen, Chien et al.'s scheme is vulnerable to a reflection attack [7] and an insider attack [32]. Ku and Chen claimed that Chien et al.'s scheme is also not reparable [32]. In addition, they also proposed an improved scheme to prevent these attacks: reflection attack and an insider attack on Chien–Jan–Tseng's scheme. In the same year, Hsu [10] pointed out that the Chien–Jan–Tseng's scheme is still vulnerable to a parallel session attack and Yoon et al. [11] claimed that the password change phase of improved scheme of Chien–Jan–Tseng's scheme is still insecure. This paper proves that security vulnerabilities still exit in Yoon et al.'s scheme is still vulnerable to parallel session attack.

*Organization*

Section II reviews Yoon et al.'s scheme [11]. Section III is about our observations on the security vulnerabilities of Yoon et al.'s scheme. Finally, comes to a conclusion in section IV.

## II. YOON ET AL.'S SCHEME

This section briefly describes Yoon et al.'s scheme [11]. This scheme has four phases: the registration phase, login phase, verification phase and the password change phase. All these four phases are described below.

### A. Registration Phase

This phase is invoked whenever $U$ initially or re-registers to *AS*. Let $n$ denotes the number of times $U$ re-registers to AS. The following steps are involved in this phase.

- ❖ User $U$ selects a random number $b$ and computes $PW_S = f(b \oplus PW)$ and submits her/his identity *ID* and $PW_S$ to the *AS* through a secure channel.
- ❖ *AS* computes two secret numbers $V = f(EID \oplus x)$ and $R = f(EID \oplus x) \oplus PW_S$, where $EID = (ID \parallel n)$ and creates an entry for the user $U$ in his account database and stores $n = 0$ for initial registration, otherwise set $n = n+1$, and $n$ denotes the present registration.

- ❖ *AS* provides a smart card to the user *U* through a secure channel. The smart card contains two secret numbers *V*, *R* and a one-way function *f*.
- ❖ User *U* enters her/his random number *b* into his smart card.

## B. Login Phase

For login, the user *U* inserts her/his smart card to the smart card reader and then keys the identity and the password to gain access services. The smart card will perform the following operations:

- ❖ Computes $C_1 = R \oplus f(b \oplus PW)$ and $C_2 = f(C_1 \oplus T_U)$. Here $T_U$ denotes the current date and time of the smart card reader.
- ❖ Sends a login request $C = (ID, C_2, T_U)$ to the *AS*.

## C. Verification Phase

Assume *AS* receives the message *C* at time $T_S$, where $T_S$ is the current date and time at *AS*. Then the *AS* takes the following actions:

- ❖ If the identity *ID* and the time $T_U$ is invalid i.e. $T_U = T_S$, then *AS* will rejects this login request.
- ❖ Checks, if $C_2 \stackrel{?}{=} f(f(EID \oplus x) \oplus T_U)$, then the *AS* accepts the login request and computes $C_3 = f(f(EID \oplus x) \oplus T_S)$. Otherwise, the login request *C* will be rejected.
- ❖ *AS* sends the pair $T_S$ and $C_3$ to the user *U* for mutual authentication.
- ❖ If the time $T_S$ is invalid i.e. $T_U = T_S$, then *U* terminates the session. Otherwise, *U* verifies the equation $C_3 \stackrel{?}{=} f(C_1 \oplus T_S)$ to authenticates *AS*.

## D. Password Change Phase

This phase is invoked whenever *U* wants to change his password *PW* with a new one, say $PW_{new}$. This phase has the following steps.

- ❖ *U* inserts her/his smart card to the smart card reader and then keys her/his identity and the old password *PW* and then requests to change the password.
- ❖ *U*'s smart cards computes $V^* = R \oplus f(b \oplus PW)$.
- ❖ Compare this calculated value $V^*$ with the secret value *V*, which is stored in the smart card of the user *U*. If they are equal, then *U* can select a new password $PW_{new}$, otherwise the smart card rejects the password change request.
- ❖ *U*'s smart cards computes a new secret number $R_{new} = V^* \oplus f(b \oplus PW_{new})$ and then replaces *R* with $R_{new}$.

## III. SECURITY ANALYSIS OF YOON ET AL.'S SCHEME

Although, Ku and Chen's scheme is modified by Yoon et al. [11] But, we analyze that Yoon et al.' scheme is still not secure. This section discusses the security weaknesses of the Yoon et al.'s scheme.

### A. Security Analysis of the Password Change Phase

This section discusses the security weaknesses of the password change of Yoon et al.'s scheme. The discussion is divided into two subsections, which are described below.

#### I. Security weaknesses in the Password Change Phase against the Outsiders

Observe the password change phase of Yoon el al.'s scheme, to replace/change the old password *PW* with a new password $PW_{new}$, the user/performer should be in possession of the old password *PW*. The following section describes how any outsider /malicious user can recover the password *PW* first and then apply this peace of information to make for the success of her/his attack.

It is clear that the smart card of a legal user *U* in Yoon et al.'s scheme contains: *the secret value V, R,* and *a random number b and a public hash function f.* According to Kocher et al. [24] and Messerges et al. [31], for the security point of view, to store the secret information in smart cards is not a good practice. On the basis of these assumptions [24,31], an antagonist is able to breach the secrets *V*, *R* and *b,* which are stored in the smart card of the user and then he will be able to perform a password guessing attack to obtain the password. For the success of this attack, by using the breached secrets *R* and *b* the adversary will perform the following operations:

- *The antagonist intercepts the login request C = (ID, $C_2$, $T_U$) and guesses a password $PW^*$.*
- *Computes $C_1^* = R \oplus f(b \oplus PW^*) = f(ID \oplus x)^*$ and $C_2^* = f(C_1^* \oplus T_U)$.*
- *Checks if $C_2^* \stackrel{?}{=} C_2$, then the adversary has correctly guessed the password $PW^*$ = PW and $C_1^* = C_1$. Otherwise, the adversary goes to step: 1.*

Once the adversary has correctly obtained $C_1$, instantly, the password $PW^*$ corresponding to $C_1$ will be the correct password and then successfully, he can change the password of the user *U*. Consequently, when the smart card was stolen, the antagonist is able to recover the password *PW* of the user and once the adversary has correctly obtain the password *PW*, then he will be able to destruct anything of his choice. Since our focus and aim is to show that the password change phase of Yoon et al.'s scheme, which is shown below that an authorized user ( antagonist) can easily replace the old password *PW*

Security Analysis of a Remote User Authentication Scheme with smart cards- Manoj Kumar

by a new password of her/his choice. For the success, the antagonist applies the following actions.

- *Inters the smart card into the smart card reader, enters the identity ID and any password PW and then requests to change the password.*
- *The smart card of the user computes $V^* = R \oplus f(b \oplus PW)$ and then compare the computed value $V^*$ with the stored value V. Obviously, both the value will be the same, because the adversary has entered the correct password. In this way, the smart card accepts the password change request.*
- *Selects a new password $PW^*_{new}$ and supplies it to the smart card reader and ultimately the smart card computes a new $R^*_{new} = R \oplus f(b \oplus PW) \oplus f(b \oplus PW^*_{new})$ and then replaces R with $R^*_{new}$.*

Thus, if the malicious user stole the user *U*'s smart card she/he will be able to make a destructive action of her/his choice. Thus, the adversary is able to change the password with a new password of his/his choice. Now the registered/ legal user *U* also will not be able to make a valid login request with her/his valid smart card because now the her/his old password *PW* will not work.

*II. Security weaknesses in the Password Change Phase against the Insider*

This subsection proves that the password change phase of Yoon et al.'s scheme is not secure against an antagonist insider at *AS*. In Yoon et al.'s scheme, observe the registration phase, the User *U* selects a random number *b* and computes $PW_S = f(b \oplus PW)$ and submits her/his identity *ID* and $PW_S$ to the *AS* through a secure channel. It means the insider of *AS* is in possession of the number $PW_S = f(b \oplus PW)$ for the legal user *U*. Again the *AS* computes two secret numbers $V = f(EID \oplus x)$ and $R = f(EID \oplus x) \oplus PW_S$, where $EID = (ID \| n)$. Thus, the insider of *AS* is also in possession of the secret numbers *V* and *R* for the legal user *U*.

Suppose the user *U* is using the same password *PW* continuously, which is supplied by the *AS* at the time of registration, then the insider at *AS* will be able to change the password *PW* with a new password of her/his choice. If the smart card is in possession of an antagonist insider at *AS* for short time, then first, the insider inters the smart card into the smart card reader and can directly supply the value *V* to the smart card reader. Either, he directly supplies *V* or in place of $f(b \oplus PW)$, he supplies the value $PW_S$ without using the hash button. Next, the antagonist insider enters a new password $PW^*_{new}$ and then the smart card computes a new $R_{new} = R \oplus f(b \oplus PW) \oplus f(b \oplus PW^*_{new})$ and then replaces R with $R_{new}$. Thus, if the malicious insider stole the user $U$'s smart card once, only for a small time and then he can replace the user's password forever in such a way that the user $U$ also will not be able to make a valid login request with her/his valid smart card because now the her/his old password $PW$ will not work properly. As a result, the Yoon et al.'s password change phase is still insecure and that is under the threat of poor reparability.

## *B. Parallel Session Attack on Yoon et al.'s Scheme*

Although, Yoon et al. [11] modified Ku and Chen's scheme to remove its security weaknesses against parallel session attack. But, we analyze that the modified scheme of Yoon et al. is still vulnerable parallel session attack. This following subsection proves our claim that the modified scheme is still vulnerable a parallel session attack by an intruder.

Since, a remote user password authentication is used to authenticate the legitimacy of the remote users over an insecure channel. Thus, an intruder Bob is able to intercept all the communication between the *AS* and user *U* and then from this intercepted information, he makes a valid login request to masquerade as a legal user. The intruder Bob applies the following steps for a successful parallel session attack on Yoon et al.'s scheme.

- ✓ Intercepts the login request $C = (ID, C_2, T_U)$ which is sent by a valid user $U$ to *AS*. In this login request $C$, the time $T_U$ is the current time of the smart card reader, whenever the user $U$ makes the login request.
- ✓ Intercepts the response message $(C_3, T_S)$, which is sent by *AS* to he user *U*. In this response message, the time $T_S$ in the current time at the *AS*, when *AS* receives the login request $C$.
- ✓ Starts a new session with the *AS* by sending a fabricated login request $C_f = (ID, C_3, T_S)$.

Upon, receiving the fabricated login request $C_f = (ID, C_3, T_S)$, at time $T_S^*$, where $T_S^*$ is the current date and time at *AS*. The *AS* performs the following steps to ensure the validity of the received login request.

- ❖ Checks the validity of the format of the identity $ID$ and the time $T_U$ i.e. $T_S^* \neq T_S$. Both these conditions hold true, because the intruder has been used a previously registered identity $ID$ and obviously the time $T_S^*$ will be different from the time $T_S$.

- ❖ Checks, the verification equation $C_3 \stackrel{?}{=} f(f(EID \oplus x) \oplus T_S)$, which is also holds truly. The logic behind the successful verification of this phase is very interesting. If we observe login and verification phase of Yoon et al.'s scheme, then it makes a sense that the second part $C_2$ of the login request $C = (ID, C_2, T_U)$ and the first part $C_3$ of the response message $(C_3, T_S)$ are computed by the same procedure and with similar information.

- ❖ $AS$ sends the pair $T_S$ and $C_3$ to the user $U$ for mutual authentication.
- ❖ If the time $T_S$ is invalid *i.e.* $T_U = T_S$, then $U$ terminates the session. Otherwise, $U$ verifies the equation $C_3 \stackrel{?}{=} f(C_1 \oplus T_S)$ to authenticate $AS$.
- ❖ Finally, $AS$ computes $C_4 = f(f(EID \oplus x) \oplus T^*_S)$ and responses with the message pair $(C_4, T^*_S)$ to the user $U$ for mutual authentication, where $T^*_S$ is the current timestamp of the $AS$. Thus, the intruder intercepts and drops this message

In this way, the fabricated login request $C_f = (ID, C_3, T_S)$, which is made by the intruder, satisfies the all the requirements for a successful authentication of the intruder Bob by the $AS$.

## IV. CONCLUSION

As, we have observed that Yoon et al. just consider the security problems in the password change phase of Ku and Chen's scheme and repaired that phase only. They again presented a modified scheme with same security parameters as it was with previous scheme. This paper analyzed that security weaknesses still exist in Yoon et al.'s scheme. The password change phase is still vulnerable to security attacks by an outsider as well as an antagonist insider at $AS$. On the other side, Yoon et al.'s scheme is still vulnerable to the parallel session attack. Thus, the security pitfalls are still exists in Yoon et al.'s scheme.

Security Analysis of a Remote User Authentication Scheme with smart cards- Manoj Kumar

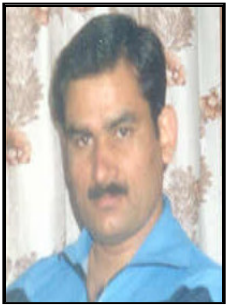

**Manoj Kumar** received the B.Sc. degree in mathematics from Meerut University Meerut, in 1993; the M. Sc. in Mathematics (Goldmedalist) from C.C.S.University Meerut, in 1995; the M.Phil. (Goldmedalist) in Cryptography, from Dr. B. R. A. University Agra, in 1996; the Ph.D. in Cryptography, in 2003. He also qualified the National Eligibility Test (NET), conducted by Council of Scientific and Industrial Research (CSIR), New Delhi- India, in 2000.

He also taught applied Mathematics at D. A. V. College, Muzaffarnagar, India from Sep 1999 to March 2001; at S.D. College of Engineering & Technology, Muzaffarnagar- U.P. – INDIA from March 2001 to Nov 2001; at Hindustan College of Science & Technology, Farah, Mathura- U.P. – INDIA, from Nov 2001 to March 2005. In 2005, the Higher Education Commission of U.P. has selected him.  Presently, he is working in Department of Mathematics, R. K. College Shamli- Muzaffarnagar- U.P. – INDIA-247776.

He is a member of Indian Mathematical Society, Indian Society of Mathematics and Mathematical Science, Ramanujan Mathematical society, and Cryptography Research Society of India. He is working as reviewer for some International peer review Journals: Journal of System and Software, Journal of Computer Security, International Journal of Network Security, The Computer Journal. He is also working a Technical Editor for some International peer review Journals- Asian Journal of Mathematics & Statistics, Asian Journal of Algebra, Trends in Applied Sciences Research, Journal of Applied Sciences. His current research interests include Cryptography and Applied Mathematics.